\documentclass{aa}  

\usepackage{graphicx,natbib,txfonts}
\def\msun{\rm M_{\odot}}
\def\kms{\rm km \, s^{-1}}

\def\lsim{\mathrel{\rlap{\lower 3pt\hbox{$\sim$}}\raise 2.0pt\hbox{$<$}}}

\def\gsim{\mathrel{\rlap{\lower 3pt\hbox{$\sim$}} \raise 2.0pt\hbox{$>$}}}

\def \mbh{$M_{\rm BH}$}
\def \Mbh{M_{\rm BH}}

\def\msunpc3{\msun~{\rm {pc^{-3}}}}

\DeclareMathOperator\atanh{atanh}

\newcommand{\be}{\begin{equation}}

\newcommand{\ee}{\end{equation}}

\def\kms{{\rm\,km\,s^{-1}}}

\begin{document}

\title{On the orientation of narrow line Seyfert I}

\author{Tullia Sbarrato\inst{1}, Massimo Dotti\inst{1,2}, Giancarlo Ghirlanda\inst{3,1} 
	    \and Fabrizio Tavecchio\inst{3}} 

\institute{Dipartimento di Fisica G.~Occhialini, Universit\`a
  		degli Studi di Milano Bicocca, Piazza della Scienza 3, I--20126 Milano, Italy
		\email{tullia.sbarrato@unimib.it}
		\and INFN, Sezione Milano-Bicocca, Piazza della Scienza 3, I--20126 Milano, Italy
		\and INAF - Osservatorio Astronomico di Brera, Merate, via E. Bianchi 46, I--23807 Merate, Italy}

\date{}

\abstract{
We study a sample of Narrow--line Seyfert 1 galaxies (NLS1) in their optical and 
radio features, in order to understand the differences between their 
radio silent, radio--loud and radio--quiet subclasses.
We first show that the different redshift and mass distributions of radio--loud and --quiet NLS1s 
could be ascribed to observational biases.
We then present a geometrical model according to which most of the different 
observational features of radio--loud and radio--quiet 
NLS1s are ascribed to the orientation of an intrinsically structured jet.
We estimate the fraction of intrinsically jetted sources among NLS1s 
that justifies the observed radio--detected population. 
Noticeably, under the assumptions of the geometrical model, 
we derive a fraction of jetted sources significantly larger than in standard AGN.
}


   \titlerunning{NLS1 orientation}
   \authorrunning{T. Sbarrato et al.}

\maketitle

\section{Introduction}

Narrow-line Seyfert 1 galaxies (NLS1s) are a sub-class of type I
active galactic nuclei (AGN) characterized by  the presence of broad emission lines (BELs) with a full
width at half maximum FWHM $\lsim 2000 \kms$ \citep{goodrich89}, weak [OIII] lines
(compared to the H$\beta$) and strong FeII emission
\citep[e.g.][]{osterbrock85, veron01}. NLS1s often show a
X-ray continuum steeper and more variable than the average AGN
\citep[e.g.][]{boller96, green93, hayashida00}.

The narrow FWHM of the optical BELs has originally been ascribed to
small BH masses, in some cases found smaller than predicted from \mbh-host
galaxy relations \citep{grupe04}.
In this ``low \mbh\ scenario'' NLS1s would be
characterized by close to Eddington accretion, with a normalized
accretion rate $f_{\rm Edd}=L/L_{\rm Edd}$ larger, on average, 
than Seyfert I with broader BELs (BLS1s) by an
order of magnitude \citep[e.g.][]{grupe04}.

Relatively narrow BELs can also be interpreted as the result of a projection
effect if the broad line region (BLR) has a disc geometry. 
Such a configuration is stable only if the gas emitting the broad lines moves
mostly in the disc plane, and, as a consequence, lines emitted by the BLR
discs that by chance are observed close to face-on would suffer a smaller Doppler
broadening.
Such dependence of the BELs FWHM on the BLR orientation has been commonly
advocated to model the properties of blazars \citep{decarli11},
radio--loud (RL) AGN \citep[e.g.][]{wills86, fine11, runnoe13}, and for quasar
in general \citep{labita06, decarli08a, shen14}. 
The BLR geometry 
can be directly tested with high quality reverberation mapping data, as the
time and frequencies at which a single broad line responds to variations
of the AGN continuum depend on the spatial distribution and velocity
field of the emitting gas \citep{blandford82}. Only recently
\cite{pancoast14b} and \cite{grier17} succeeded in constraining the disc
geometry of a few BLRs taking advantage of high quality reverberation
data coupled with the parametric dynamical model proposed by
\cite{pancoast14a}. They modeled the individual BLRs as a collection of
clouds whose positions and velocities are extracted by parametric
distributions. In particular, the spatial distribution of the clouds would
depend on the geometrical thickness of the BLR $\theta_0$, ranging from
0$^{\circ}$ for a thin disc to $90^{\circ}$ for a spherical distribution. 
A nested sampling search through the parameter space resulted in 8 BLR
(out of 9 AGN studied by the two teams) described by discs with thickness 
$\theta_0 < 40^{\circ}$ and inclination with respect to the line of sight
$\theta_i< 40^{\circ}$.

\cite{decarli08b} proposed a ``geometrical scenario'' for NLS1s assuming
that these are standard Seyfert Is hosting BLRs with disc geometries
observed close to face-on \citep[see also][]{bian04b}.
The angle within
which AGN are seen as NLS1 has hence been estimated from their relative
abundance. The resulting correction factor to \mbh\ obtained from the
de-projection shifts the distributions of NLS1 masses and $f_{\rm Edd}$ on the
top of those obtained from BLS1s, in agreement with independent estimates of
the NLS1 \mbh\ \citep{calderone13}. 
Such a scenario could in principle be
tested by reverberation mapping studies. Unfortunately, only 3 objects in
\cite{pancoast14b} and \cite{grier17} are NLS1s, one of which (Mrk
1310) have been modeled as a disc with $\theta_i \approx 8^{\circ}$, in
agreement with the geometrical scenario, while the other 2 (Mrk 335 and PG
2130+099) are best modeled by higher inclinations ($\theta_i\sim 30^{\circ}$ in
both cases).

The recent developments in the study of RL-NLS1s and
the detection of $\gamma$ emission from NLS1s by the {\it Fermi} satellite
\citep[e.g.][]{abdo09, foschini11, foschini12, dammando12, dammando15} has re-ignited
the interest in the field. The $\gamma$ emission is caused by a
relativistic jet quasi-aligned with the line of sight. 
Indeed RL-NLS1s have been interpreted as an  early evolutionary 
phase that would eventually lead to fully developed blazars \citep{foschini15, foschini17}.
It is interesting to note that the radio-to-optical
spectral energy distribution (SED) and variability of a large fraction of RL-NLS1s
seem to indicate that the low frequency emission is
dominated by a relativistic jet pointing toward us
\citep{lahteenmaki17}.

In the few NLS1s where the jet orientation was derived, 
the observed degree of jet alignment supports the geometrical
scenario\footnote{In the RL-NLS1 PKS 2004-447 such scenario as been
  successfully tested through the spectro-polarimetry of the broad H$\alpha$
  line \citep{baldi16}.}.  
Under the geometric assumption, 
RL-, radio--quiet (RQ) and radio silent NLS1 should differ
for the presence of a jet and its detectability.
Such assumption has been debated as the mass
and redshift distributions of RL-NLS1s and RQ-NLS1s seem to differ
\citep{jarvela15}. 
A tentative comparison between RL- and RQ-NLS1 populations was also
performed by \cite{berton15}, in their attempt to identify a parent population 
of the flat spectrum radio--loud NLS1s among steep spectrum RL-NLS1s, 
RQNLS1s and disk--hosted radio galaxies. Unfortunately, the flat spectrum RL- 
and RQ-NLS1 samples are still too poorly populated to draw strong conclusions. 

In this study we first highlight some observational biases
that could account for such apparent discrepancies
(section~\ref{sec:distinction}), then we provide a physical
interpretation of the observed $\gamma$-NLS1s and RL-NLS1s in the context of
a simple model built on the assumptions of the ``geometrical scenario''
(section~\ref{sec:frec}). Possible implications of the ``geometrical
scenario'' are discussed in section~\ref{sec:disc}.

\section{Do RL-NLSy1 and RQ-NLSy1 belong to two distinct populations?}
\label{sec:distinction}

\begin{figure}
\includegraphics[trim=30 180 30 200, clip, width=0.5\textwidth]{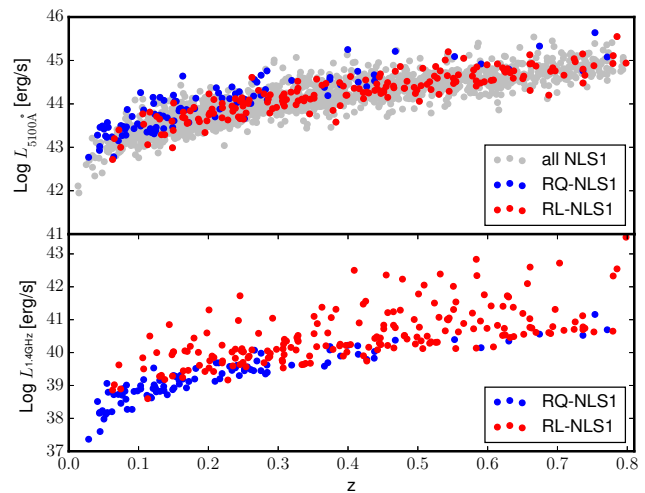}
\vskip -0.2cm
\caption{Distribution of 5100\AA\ ({\it top}) and radio ({\it bottom}) luminosities vs.\ redshift for the NLS1
  subclasses. RQ-NLS1 (blue dots) are mainly located at low $z$.  The
  few sources at higher $z$ are among the most luminous: the radio
  flux sensitivity only allows to observe the peak of the iceberg.
  RL-NLS1 (red dots), instead, are more evenly distributed in
  luminosities and $z$: thanks to their larger radio flux, they are 
  less affected by this observational bias compared with RQ sources. 
  They will be missing in radio surveys from higher redshift than they RQ counterparts.  
  }
\label{fig:L5100}
\end{figure}

\cite{jarvela15} studied the sources in \cite{zhou06} sample
of NLS1 galaxies from the Sloan Digital Sky Survey (SDSS), 
by classifying them on the basis of their radio emissions 
and studying the distribution of the sub--populations. 
Starting from a sample of 2011 sources in \cite{zhou06}, 
\cite{jarvela15} classified the sources in  
radio undetected NLS1s, RQ-NLS1s detected by the FIRST
survey (97 sources, with a radio--loudness parameter defined as the ratio of 1.4
GHz radio flux density over 400 nm optical flux density R$<$10), and
RL-NLS1s (195 sources with R$>$10). 
\cite{jarvela15} found different redshift
distributions for RQ-NLS1s and RL-NLS1s (mean $z=0.22$ and $z=0.41$
respectively), concluding that the two classes belong to intrinsically
different populations. Most of radio--detected sources are
close to the FIRST flux limit (i.e.\ 1mJy), possibly
introducing some observational bias.
While RL radio emission is dominated by
the jet, detected RQ sources are often dominated by star
formation in the same bands \citep{caccianiga15,jarvela15, padovani16},
often not showing jet features at all. Being intrinsically fainter in radio,
RQ-NLS1s are therefore biased toward lower $z$, while RL sources
can be detected up to higher $z$.  
This can be inferred from Fig.\ \ref{fig:L5100}, where we show 
the optical and radio luminosities ($L_{5100}$, $L_{\rm 1.4GHz}$) 
vs.\ the redshifts of \cite{jarvela15} NLS1 samples.  
RL-NLS1 sources are more uniformly distributed in both planes, as
the non radio--detected sources.  
RQ-NLS1, instead, are mainly located at low $z$. 
We stress that less luminous but intrinsically RL-NLS1s could be 
classified as radio-silent, if their radio fluxes are below the FIRST 
detection limit, affecting the population studies of jetted NLS1. 
Some of these, in fact, could be classified as radio-silent sources, 
preventing us from distinguishing between intrinsically jetted and non-jetted sources.

In addition to the properties directly measured from the observations,
\cite{jarvela15} provided virial estimates of the \mbh\ for
each object following \cite{greene05}:
\be\label{eq:SE}
\Mbh \propto L_{5100}^{0.64} \times {\rm FWHM}({\rm H}\beta)^2, 
\ee
where FWHM is the full width at half maximum of the broad H$\beta$
line\footnote{Such estimate does not take into account the effect of
  the individual BLR orientation, as discussed in the following
  section.} and $L_{5100}$ is the monochromatic luminosity at 5100\AA. 
  \cite{jarvela15} concluded that the RQ- and RL-NLS1
populations have different \mbh\ distributions (with mean
$\log($\mbh$/M_\odot)=6.96$ and 7.18, respectively). The bias toward
higher redshifts for RL-NLS1s can be (at least partially) responsible
for the different \mbh\ distributions. 
A second possible bias affecting the RL-NLS1s \mbh\ is related to 
the non--thermal jet emission itself. 
As explored by \cite{lahteenmaki17}, when observed at multiple radio
frequencies (1.4, 37GHz), RL-NLS1 often show beaming and variability
features typical of blazars, i.e.\ sources with jets directed toward our line of sight.  
In that case, the jet emission often contributes to optical emission \citep{ghisellini17, giommi12}, 
resulting in an overestimate of the 5100\AA\ continuum luminosity
henceforth of  \mbh.
A significant contribution from the jet to the optical luminosity of
RL-NLS1s has already been discussed in literature. \cite{calderone12} studied
the example of B2 0954+25A, a blazar that on different observations
taken at different epochs shows an optical emission either dominated
by the jet or by the accretion disc\footnote{Similar results have been
presented for NLS1 PKS 2004-447 by \cite{abdo09} and for PMN J0948+0022,
SBS 0846+513, PMN J0948+0022, 1H 0323+342 and PKS 1502+036 in
\cite{dammando16}}. Jets have been demonstrated to play an important
role in the optical variability of RL-NLS1s in the Catalina Real Time
Transient Survey \citep{rakshit17}. 
Note that a substantial jet contamination in the optical luminosity could affect 
the observed NLS1 spectra: the equivalent widths of broad emission lines 
in jet--dominated sources should be smaller than their non--jetted counterparts. 
This could serve as a possible diagnostic to understand whether RL-NLS1s are 
truly contaminated by jet emission in their optical luminosity.

\begin{figure}
\includegraphics[trim=30 180 30 200, clip, width=0.5\textwidth]{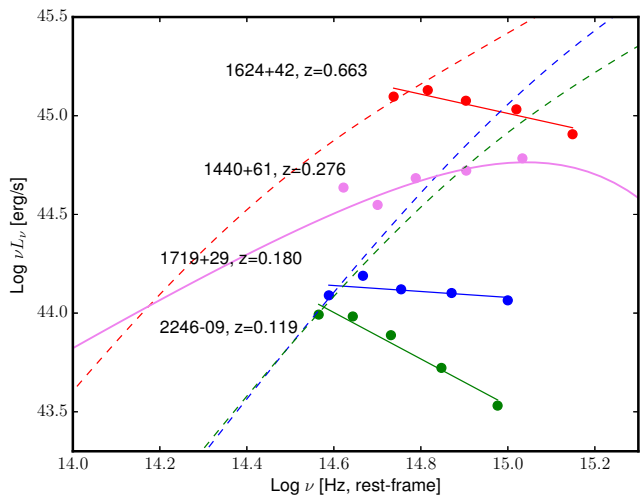}
\vskip -0.2cm
\caption{Optical SEDs for three 
RL-NLS1s and a RQ-NLS1. 
Green, red and blue data show the decreasing profiles of 
SDSS J224605.44-091925.1 ($z=0.119$), SDSS J162458.42+423107.5 
($z=0.663$) and SDSS J171930.56+293412.8 ($z=0.180$), 
typical of jet--dominated optical emissions.
The analogous dashed lines show \cite{ss73} disc profiles for the three sources, 
drawn with the virial masses by \cite{jarvela15} and normalised by 
forcing them to pass through the lowest frequency data point. 
Violet data, instead, show the disc--dominated profile of 
the RQ-NLS1 SDSS J144012.76+615633.2 ($z=0.276$).
The RL-NLS1s have an SED clearly not disc--like.
}
\label{fig:slope}
\end{figure}

To test whether a jet--like
power--law could contribute to the optical continuum in J\"arvela et
al.\ RL-NLS1s, we studied their optical--UV slopes.  At such frequencies, 
the jet emission is characterised by a negative power--law in
$\nu-\nu L_\nu$ \citep[see e.g.][]{ghisellini17}, differently from the big blue
bump typical of the disc emission.
Hence, we studied the RL-NLS1 SEDs 
in the SDSS band. 
We noted that 152 out of 195 RL-NLS1s show a negative slope 
among the SDSS bands, with a profile
similar to the blue data in Figure \ref{fig:slope}.
Of these, 23 out of 195 sources have a monotonically decreasing 
$\nu L_\nu$ or show a single fluctuation from the monotonic profile, 
as the red and green examples in Figure \ref{fig:slope}.
We are aware that this characterisation does not guarantee a
good sampling of jet--dominated sources, and a
broader--band SED analysis would be needed to strongly confirm it.
Nonetheless, this points toward a partial continuum contamination by
the jet in a large number of RL-NLS1s, implying an overestimate of 
$L_{5100}$, and therefore of \mbh\ compared to RQ-NLS1s.
Another possible cause for such a negative slope could be intrinsic dust absorption. 
Figure \ref{fig:slope} includes three dashed lines, 
each representing a hypothetical 
accretion disc, modelled as a \cite{ss73} multicolor black body, 
with virial masses derived by \cite{jarvela15}, 
and accretion rates set by fitting the lowest frequency point to the 
continuum disc emission of each object with decreasing slopes.  
If reddening by absorption had an important role for these sources, 
these disc profiles would be the lowest accretion rate solutions 
possible. By forcing this ''fit'', we obtain Eddington ratios of 
$L_{\rm d}/L_{\rm Edd}\simeq100$, 10 and 420 for green, red and blue data respectively. 
A standard accretion disc with a fair amount of ionising radiation is 
not stable at such high Eddington ratios. 
Dust absorption has most likely a minor role in shaping 
the UV-optical SEDs of these sources.

A definite interpretation of the relative jet and disc contributions in the optical,
along with a 
demonstration that the two distributions may be extracted from the
same population,  would require a detailed SED modeling including the 
contribution of all the components of an AGN.
We are
currently working on implementing the required modifications in
QSFIT\footnote{http://qsfit.inaf.it/}, a free software package to
automatically perform the analysis of AGN spectra
\citep{calderone16}. We postpone such analysis to a future
investigation, while here we simply assume that orientation is the
main parameter determining the classification of AGN as Seyfert I and
NLS1s, and, for radio detected sources, 
between RQ-NLS1s, RL-NLS1s and
$\gamma$-NLS1s. The consequences of such model assumption are
highlighted in the next section.

\section{Geometrical unification model for NLS1s}
\label{sec:frec}

In the geometrical scenario the critical angle that discriminates
between NLS1s and BLS1s ($\theta_{\rm NLS1}$) can be obtained from the
number ratio of the members of the two populations
\citep{decarli08b}. They found $13^{\circ}\lsim \theta_{\rm NLS1}
\lsim 19^{\circ}$.  Here we extend their model to the sub-sample of
RL-NLS1s and $\gamma$-NLS1s, under the assumptions that: ($i$) only a
fraction $f_{\rm jet}$ of NLS1s possess a jet, and ($ii$) that only
objects that are seen close to face--on would be observed in $\gamma$
or selected as RL (within a maximum inclination of
$\theta_{\gamma-{\rm NLS1}}$ and $\theta_{\rm RL-NLS1}$
respectively)\footnote{This last assumption is somewhat extreme, as an
  AGN with bright radio--lobes could be RL even if the radio core
  emission is not detected. However, most of the RL-NLS1s show clear
  signatures of a blazar nature and are close to the radio
  detection limit.}.
The second assumption is motivated by
the ``spine-layer'' (SL) jet model, in which the central and fastest
part of the jet, responsible for the $\gamma$ emission, is embedded in
a slower layer contributing to the radio flux \citep{ghisellini05}.
As reference values $\theta_{\rm RL-NLS1}\approx 7^{\circ}$ and
$\theta_{\gamma-{\rm NLS1}}\approx 4^{\circ}$, 
reasonable in the SL framework that we are following.

By simply comparing the solid angles involved, we can correlate the
two critical angles with the ratio between the number of RL-NLS1s
($N_{\rm RL-NLS1}$) and that of all the NLS1 ($N_{\rm NLS1}$):
\be\label{eq:RL} \frac{N_{\rm RL-NLS1}}{N_{\rm NLS1}} =
\frac{(1-\cos(\theta_{\rm RL-NLS1}))}{(1-\cos(\theta_{\rm NLS1}))}
\times f_{\rm jet}, \ee where $f_{\rm jet}$ corrects for the intrinsic
fraction of jetted sources.
Under the assumption of a negligible radio flux limit effect, we can 
estimate $f_{\rm jet}$ as the
fraction of RL-NLS1 $N_{\rm RL-NLS1}/N_{\rm NLS1}=195/2011$ observed
by \cite{jarvela15}, we obtain $0.3 \lsim f_{\rm jet} \lsim
0.7$, where the scatter depends on the assumed value of $\theta_{\rm
  NLS1}$. Interestingly, the fraction of intrinsically jetted NLS1s is
quite higher than what typically assumed for AGN. This is not
unexpected, though, as the high degree of alignment of the jet with
the line-of-sight allows for the observation of intrinsically faint
radio sources.

\begin{figure}
\includegraphics[width=0.45\textwidth]{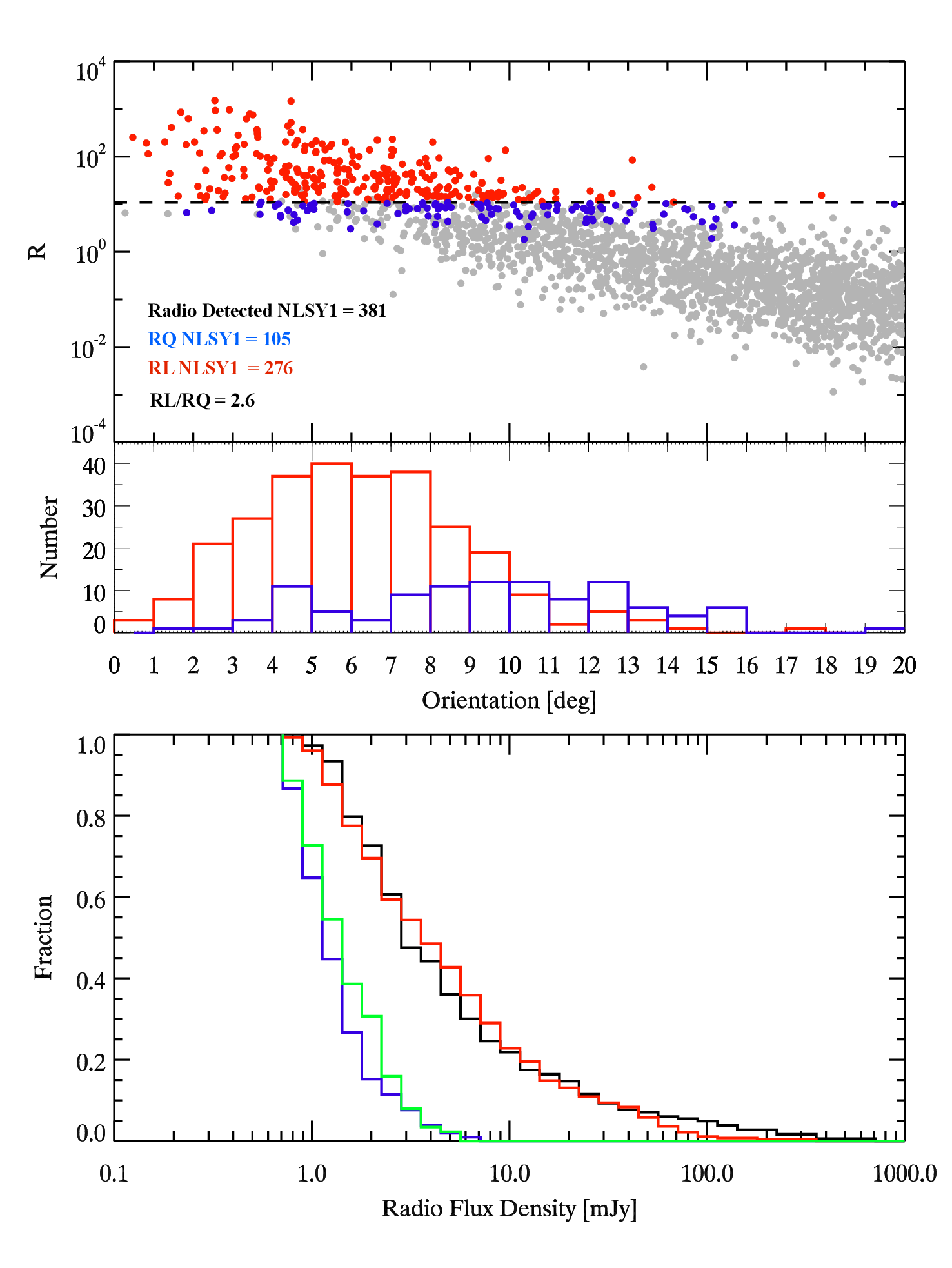}
\vskip -0.2cm
\caption{
{\it Upper panel:} radio--loudness as a function of the orientation angle for the 
simulated NLS1 sample. 
{\it Middle panel:} orientation angle distribution for simulated radio--detected NLS1s.
Our simulated samples have average orientations of 
$(6.22\pm2.72)^\circ$ (RL-NLS1) and $(9.56\pm3.53)^\circ$ (RQ-NLS1), 
and we can infer $\theta_{\rm RL-NLS1}\sim9^\circ$.
{\it Bottom panel:} radio flux density distributions for simulated and real samples.
Red and blue symbols and lines in all panels refer to simulated RL- and RQ-NLS1s, while
black and green lines in the bottom one refers to the observed RL and RQ populations 
present in J\"arvela et al.\ (2015).
}
\label{fig:mc}
\end{figure}

As a consistency check 
that accounts for the radio flux limit effect,
we performed a Monte-Carlo simulation 
to reproduce the radio--detected NLS1 sample by \cite{jarvela15}.
We start from the NLS1 sample by \cite{zhou06} and randomly select 
a source with its $z$ and observed optical luminosity. 
To this source, we assign (i) a random jet bulk velocity $\Gamma$ 
and (ii) an intrinsic comoving radio luminosity $L^\prime_{\rm r}$, 
both extracted from Gaussian distributions. 
Furthermore, we assigned a jet orientation $\theta_{\rm v}$ (with respect to the observer 
line of sight)  drawn from a solid angle weighted 
distribution within a maximum angle of 20$^\circ$.
With these parameters, we computed the source radio observed flux:
\begin{equation}
F_{\rm r} = \frac{L^\prime_{\rm r}\delta^4}{4\pi D_{\rm L}(z)^2}
\label{eq:beam}
\end{equation}
where $\delta=1/\Gamma(1-\beta\cos\theta_{\rm v})$ is the jet beaming factor 
and $D_{\rm L}(z)$ is the luminosity distance.
For each source, we thus compute the radio--loudness $R=F_{\rm r}/F_{\rm opt}$ 
where for $F_{\rm opt}$ we used the observed optical flux of this source 
as reported in \cite{zhou06} catalog. 
Our aim is to reproduce the radio--loudness properties of 
\cite{jarvela15} sample of NLS1, in terms of R and radio flux distributions, total number of 
radio detected RL- and RQ-NLS1s, and the relative number of RL- and RQ-NLS1s 
(these are our constraints). 
Therefore, we select in our simulated sample only those source with 
a radio flux larger than FIRST radio flux limit (1mJy) to be compared 
with \cite{jarvela15} sample. 
The free parameters of this Monte Carlo test are the central values 
and width of $\Gamma$ and $L^\prime_{\rm r}$ distributions. 
Since this is a consistency check, we do not aim at fully exploring 
their parameter space, but only testing if reasonable assumptions on their values 
can reproduce our constraints. 
Fig.\ \ref{fig:mc} shows the radio--loudness distribution as a function 
of the orientation in our simulated sample (upper panel), obtained by 
sampling from distributions of $\Gamma$ centred at 10 ($\sigma=0.2$) and 
of $L^\prime_{\rm r}$ centred at $8\times10^{35}$ erg s$^{-1}$ ($\sigma=0.2$dex).
The simulation needs $f_{\rm jet}\sim0.5$ to reproduce the RL occurrence in the sample\footnote{ 
The fraction of intrinsically jetted sources increases up to 1 by increasing the maximum 
viewing angle up to $\sim30^\circ$ 
in order to obtain a simulated sample consistent with the one by \cite{jarvela15}. 
}, 
and suggests an average orientation angle of $(6.22\pm2.72)^\circ$ for RL-NLS1s,
roughly consistent with our assumption on $\theta_{\rm RL-NLS1}$ ($7^\circ$). 
Interestingly, our test suggests that $\sim1/3$ RQ-NLS1s are 
dominated by SF in their radio emission, since they 
result underestimated in proportion to RL-NLS1s, if only the jet emission 
is considered contributing to the radio luminosity 
\citep[in agreement with][]{jarvela15}.
As a note of caution, we stress that this test does not aim at reproducing source by source 
the NLS1s in \cite{jarvela15}, but rather their distribution 
(of R, radio fluxes and RL/RQ) as a radio flux--limited statistical sample. 
For completeness, we assumed an intrinsic optical--to--radio luminosity ratio 
for the jet component as 
in the jet--dominated state of B2 0954+25A \citep{calderone12}. 
We similarly beamed the optical luminosity, derived the optical flux 
(see Eq.\ \ref{eq:beam}), and compared it with the observed optical flux. 
We find that in all the RQ sources the simulated optical flux is on average 
1/50-1/100 of the observed one. In RL sources, instead, a minor 
fraction ($\sim5\%$) overestimates the observed optical flux, while in the 
majority of RL sources the jet emission likely contributes on the same level as the disc emission, 
being comparable but less than the observed flux. 
This effect, coupled with the redshift bias introduced in \S\ref{sec:distinction}, 
is likely responsible of the apparently different black hole mass distributions in the 
RQ- and RL-NLS1 subclasses.

The previous analytic argument applied to the fraction of RL-NLS1
can be used to constrain the fraction of RL-NLS1s
that would be observed with a $\gamma$-ray selection:
\be\label{eq:gamma} \frac{N_{\gamma-{\rm NLS1}}}{N_{\rm RL-NLS1}} =
\frac{(1-\cos(\theta_{\gamma-{\rm NLS1}}))}{(1-\cos(\theta_{\rm
    RL-NLS1}))} \times \eta_{Fermi}. \ee In this case, we will assume
that all the RL-NLS1s host a jet (i.e. the contamination by SF
radio emission is negligible), so that 
$\eta_{Fermi}$ accounts for the
possible selection effects introduced by the {\it Fermi}/LAT flux limit. 
The ratio $N_{\gamma-{\rm
    NLS1}}/N_{\rm RL-NLS1}=1/7$ observed by \cite{dammando15} results
in $\eta_{Fermi}\approx 0.43$, i.e. about half of the NLS1s with
very aligned jets have been observed by {\it Fermi}.

The critical angles for a jetted NLS1 to be classified as RL-NLS1 or
$\gamma$-NLS1 can be used to improve the estimates of their
masses. Under the assumption that the BLR is virialized and lies
within the influence sphere of the BH, \mbh\ can be estimated through:
\be\label{eq_virial} M_{\rm BH} \approx \frac{R_{\rm BLR} v_{\rm
    BLR}^2}{G}, \ee where $R_{\rm BLR}$ is an estimate of the broad
line region (BLR) size, and $v_{\rm BLR}$ the typical velocity of BLR
clouds.  Eq.~\ref{eq_virial} implies Eq.~\ref{eq:SE} when $R_{\rm
  BLR}$ is estimated through the $R_{\rm BLR}$--luminosity relations
\citep{kaspi00, kaspi05, bentz06, bentz09b} and $v_{\rm BLR}$ is
inferred from the H$\beta$ width $v_{\rm BLR} = f \cdot {\rm FWHM}$.
Here $f$ is a fudge factor that depends on the geometry of the BLR
that is usually set by comparing the \mbh\ estimates obtained from
Eq.~\ref{eq:SE} with independent mass estimates over a sample of Type
I objects.

Under the assumption of a thick disc BLR \citep[e.g.][]{collin06}, $f$
can be expressed for any  AGN as: \be\label{eq_f} f=\left[2
  \sqrt{\left(\frac{H}{R}\right)^2+\sin^2{\theta_i}}\right]^{-1}, \ee
where $\theta_i$ is the angle between the normal to the BLR disc and
the line of sight, and the aspect ration $H/R$ is related to the
relative importance of isotropic (e.g.  turbulent) versus rotational
motions. Here we will assume $H/R \approx 0.1$, since such value can
resolve the possibly not physical discrepancy between NLS1 and BLS1
black hole masses \citep{decarli08b}.
Each class of AGN will have a different bias in the \mbh\ estimates,
proportional to the average value of $f^2$ ($\overline{f^2}$) over all the allowed
orientations:
\begin{equation}
  \frac{\int^{\theta_{\rm M}}_0 f^2 {\rm d}\Omega }{\int^{\theta_{\rm M}}_0{\rm d}\Omega}= \frac{1}{4\sqrt{(H/R)^2+1}}
  \frac{\atanh\left(\frac{\cos{\theta_i}}{\sqrt{(H/R)^2+1}}\right)\biggr|^{\theta_{\rm M}}_{0}}{1-\cos{\theta_{\rm M}}},
\end{equation}
where $\theta_{\rm M}$ is equal to $\theta_{\rm BLS1}=40^{\circ}$
for the whole Seyfert I population, $\theta_{\rm NLS1}=15^{\circ}$ for
NLS1s, $\theta_{\rm RL-NLS1}=7^{\circ}$ for RL--NLS1s and
$\theta_{\gamma-{\rm NLS1}}=4^{\circ}$ for $\gamma$-NLS1s.
The correction factor to remove the projection effects for a specific
class of AGN is therefore $\alpha_{\rm class}=\overline{f^2_{\rm
    class}}/\overline{f^2_{\rm BLS1}}$ with $\alpha_{\rm NLS1}\approx
6.3$, $\alpha_{\rm RL-NLS1}\approx 12.8$, and $\alpha_{\gamma-{\rm
    NLS1}}\approx 17.1$.
We stress that: (1) the larger correction for the RL-NLS1s would
increase the disagreement between the observed distributions of
\mbh\ in RQ-NLS1s and RL-NLS1s discussed by \cite{jarvela15}, hinting
at an even larger impact of the observational biases discussed in
section~\ref{sec:distinction}; (2) the limited variation between
$\alpha_{\rm RL-NLS1}$ and $\alpha_{\gamma-{\rm NLS1}}$ is due to the
$H/R$ term in Eq.~\ref{eq_f}. A larger value of $H/R\approx 0.2$ would
result in smaller and more similar correction factors: $\alpha_{\rm
  NLS1}\approx 3.6$, $\alpha_{\rm RL-NLS1}\approx 5.2$, and
$\alpha_{\gamma-{\rm NLS1}}\approx 5.8$.

\section{Discussion}
\label{sec:disc}

We presented a simple geometrical unification model based on the assumption
that a fraction of RQ-NLS1s ($\sim50\%$), 
RL-NLS1s and $\gamma$-NLS1s are classified as such only because
of their orientation.  Here we discuss our model in the broader context of
NLS1s, commenting on possible tests.

Within the same scenario, \cite{smith04, smith05} have theoretically
predicted that, when observed in polarized light, NLS1s should have a
low polarization fraction, the polarization angle should change
monotonically from one wing of the BELs to the other, and the BELs
should be significantly broader than those observed in direct light
\citep{smith04, smith05}. Here it is assumed that polarized BELs are
scattered into the line of sight by material that is close to coplanar
with the BLR (i.e. the torus).  Such prediction has been searched for
in RQ-NLS1s, but many objects have not been detected in polarized
light, and none of the detected show any significant
broadening of the BELs \citep[e.g.][]{goodrich89, breeveld98,
  kay99}. On the other hand, only one RL-NLS1 (PKS 2004-447) has been
object of a spectropolarimetry study \citep{baldi16}, and the three
predicted features have been observed, with a FWHM in the polarized
H$\alpha$ of $\approx 9000 \kms$. Here we propose for the first time a
simple scenario that can resolve such apparent tension between the two
classes of NLS1s. We speculate that in 
those NLS1s that are intrinsically without a jet (a fraction of the RQ-NLS1)
a significant fraction of polarized light could be due to
polar scattering of the BELs, by material above the BLR, as observed
for type II AGN \citep{antonucci83, miller83, antonucci85}. A
polar-scattered component could easily overwhelm the planar-scattering
contribution, predicted to result in low polarization fractions for
face-on objects because of the axi-symmetry of the scattering
material. In RL-NLS1s, on the contrary, the jet could evacuate the
intervening material, resulting in a lower degree of polarization in
which only the broadened component of the BEL is left. The evacuation
of the polar material by a jet closely aligned with the line of sight
has been already proposed by \cite{ghisellini16} to account for the
lack of high-$z$ blazar parent population. Such simple speculative
scenario can be tested through spectropolarimetry of a larger sample
of RL-NLS1s.

As a note of caution, we discussed only the radio and $\gamma$ features
of NLS1s, and the properties of their optical BELs. Other peculiarities are
characteristic of this class of AGN. As an example, NLS1s show weak [OIII]
emission when normalized to the H$\beta$ line, explained both by the
low-\mbh\ and geometrical scenarios as a consequence of weak thermal
emission at high (UV) frequencies. While in the low-\mbh\ scenario this high
energy cut-off is supposedly caused by the self shielding of the optically
and geometrically thick innermost (hotter) parts of a close-to- or
super-Eddington accretion disc \citep[see e.g.][]{boroson02}, in the
geometrical scenario such self-shielding is not required, as the innermost
parts of the accretion discs are naturally colder. We stress that such
geometrical model is in agreement with the infrared-to-UV SED fitting
discussed for example by \cite{calderone13}. In addition, NLS1s show a
steeper/more variable X-ray continuum with respect to the bulk of the type I
AGN population. These features have been interpreted as due to higher
Eddington ratios in the low-\mbh\ scenario\footnote{Although, a SED fitting
  including the soft X-ray part of the NLS1s spectra seems to indicate MBH
  masses larger than those obtained from the virial estimates and, as a
  consequence, smaller inclinations \citep[within $\sim
    17^{\circ}$,][]{bian04b}}, but is also consistent with the ``aborted
jet'' model \citep{ghisellini04}, in which a weak radio jet (similar to the
ones discussed in the geometrical model for RL- and $\gamma$-NLS1s) is
responsible for the highly variable X-ray emission. In both models the
Comptonization of optical-UV photons produced in the accretion disc onto the
non-thermal electrons is responsible for the steepness of the X-ray
spectrum, as long as most of the accretion energy is released in the
accretion disc and not directly into the non-thermal electrons of the
corona/jet. Although the geometrical model explains many
different aspects of the NLS1s and in particular jetted NLS1s, further
theoretical effort is needed to reproduce all the features that characterize
such class of AGN. In particular, to our knowledge the peculiar FeII
intensity is not an immediate prediction of the geometrical model, and
certainly requires some additional investigation.

In conclusion, the geometrical model has already been successfully applied to 
a RL-NLS1 \citep{baldi16}, and it could explain the blazar-like behavior of the
RL sub-class.
Nevertheless we cannot firmly constrain the
relative fraction of face--on NLS1s vs. those powered
by an undermassive BH accreting close to its Eddington limit. Of
particular interest in this regard is the study by \cite{sani10}, in
which the ratio between star formation activity and AGN
emission ($R$) has been estimated both for NLS1s and BLS1s. The $R$
parameter does not depend in any way on the assumed mass of the
central BH, hence is not affected by possible orientation
effects. Although a significant fraction of NLS1s lies in the same
region of the (\mbh-$R$) plane as the BLS1s, the NLS1 sample shows
overall an enhanced star formation with respect to the BLS1s.  It
would be interesting to check whether the star formation properties
can be used to discriminate among the two scenarios.

\begin{acknowledgements}
We thank Luigi Foschini and Valentina Braito for useful discussions. 
\end{acknowledgements}

\bibliographystyle{aa}
\bibliography{NLS1bis}

\begin{thebibliography}{60}
\expandafter\ifx\csname natexlab\endcsname\relax\def\natexlab#1{#1}\fi

\bibitem[{{Abdo} {et~al.}(2009){Abdo}, {Ackermann}, {Ajello}, {Baldini},
  {Ballet}, {Barbiellini}, {Bastieri}, {Bechtol}, {Bellazzini}, {Berenji},
  {Bloom}, {Bonamente}, {Borgland}, {Bregeon}, {Brez}, {Brigida}, {Bruel},
  {Burnett}, {Caliandro}, {Cameron}, {Caraveo}, {Casandjian}, {Cecchi}, {{\c
  C}elik}, {Chekhtman}, {Cheung}, {Chiang}, {Ciprini}, {Claus}, {Cohen-Tanugi},
  {Conrad}, {Cutini}, {Dermer}, {de Palma}, {Silva}, {Drell}, {Dubois},
  {Dumora}, {Farnier}, {Favuzzi}, {Fegan}, {Focke}, {Foschini}, {Frailis},
  {Fukazawa}, {Fusco}, {Gargano}, {Gehrels}, {Germani}, {Giebels}, {Giglietto},
  {Giordano}, {Giroletti}, {Glanzman}, {Godfrey}, {Grenier}, {Grove},
  {Guillemot}, {Guiriec}, {Hayashida}, {Hays}, {Horan}, {Hughes},
  {J{\'o}hannesson}, {Johnson}, {Johnson}, {Kadler}, {Kamae}, {Katagiri},
  {Kataoka}, {Kerr}, {Kn{\"o}dlseder}, {Kuss}, {Lande}, {Latronico}, {Longo},
  {Loparco}, {Lott}, {Lovellette}, {Lubrano}, {Makeev}, {Mazziotta},
  {McConville}, {McEnery}, {Meurer}, {Michelson}, {Mitthumsiri}, {Mizuno},
  {Monte}, {Monzani}, {Morselli}, {Moskalenko}, {Murgia}, {Nolan}, {Norris},
  {Nuss}, {Ohsugi}, {Omodei}, {Orlando}, {Ormes}, {Pelassa}, {Pepe}, {Persic},
  {Pesce-Rollins}, {Piron}, {Porter}, {Rain{\`o}}, {Rando}, {Razzano},
  {Rochester}, {Rodriguez}, {Ryde}, {Sadrozinski}, {Sambruna}, {Sander}, {Saz
  Parkinson}, {Scargle}, {Sgr{\`o}}, {Smith}, {Spandre}, {Spinelli},
  {Strickman}, {Suson}, {Tagliaferri}, {Takahashi}, {Takahashi}, {Tanaka},
  {Thayer}, {Thayer}, {Thompson}, {Tibaldo}, {Tibolla}, {Torres}, {Tosti},
  {Tramacere}, {Uchiyama}, {Usher}, {Vasileiou}, {Vilchez}, {Vitale}, {Waite},
  {Wang}, {Winer}, {Wood}, {Ylinen}, {Ziegler}, {Fermi/LAT Collaboration},
  {Ghisellini}, {Maraschi}, \& {Tavecchio}}]{abdo09}
{Abdo}, A.~A., {Ackermann}, M., {Ajello}, M., {et~al.} 2009, \apjl, 707, L142

\bibitem[{{Antonucci}(1983)}]{antonucci83}
{Antonucci}, R.~R.~J. 1983, \nat, 303, 158

\bibitem[{{Antonucci} \& {Miller}(1985)}]{antonucci85}
{Antonucci}, R.~R.~J. \& {Miller}, J.~S. 1985, \apj, 297, 621

\bibitem[{{Baldi} {et~al.}(2016){Baldi}, {Capetti}, {Robinson}, {Laor}, \&
  {Behar}}]{baldi16}
{Baldi}, R.~D., {Capetti}, A., {Robinson}, A., {Laor}, A., \& {Behar}, E. 2016,
  \mnras, 458, L69

\bibitem[{{Bentz} {et~al.}(2009){Bentz}, {Peterson}, {Netzer}, {Pogge}, \&
  {Vestergaard}}]{bentz09b}
{Bentz}, M.~C., {Peterson}, B.~M., {Netzer}, H., {Pogge}, R.~W., \&
  {Vestergaard}, M. 2009, \apj, 697, 160

\bibitem[{{Bentz} {et~al.}(2006){Bentz}, {Peterson}, {Pogge}, {Vestergaard}, \&
  {Onken}}]{bentz06}
{Bentz}, M.~C., {Peterson}, B.~M., {Pogge}, R.~W., {Vestergaard}, M., \&
  {Onken}, C.~A. 2006, \apj, 644, 133

\bibitem[{{Berton} {et~al.}(2015){Berton}, {Foschini}, {Ciroi}, {Cracco}, {La
  Mura}, {Lister}, {Mathur}, {Peterson}, {Richards}, \& {Rafanelli}}]{berton15}
{Berton}, M., {Foschini}, L., {Ciroi}, S., {et~al.} 2015, \aap, 578, A28

\bibitem[{{Bian} \& {Zhao}(2004)}]{bian04b}
{Bian}, W. \& {Zhao}, Y. 2004, \mnras, 352, 823

\bibitem[{{Blandford} \& {McKee}(1982)}]{blandford82}
{Blandford}, R.~D. \& {McKee}, C.~F. 1982, \apj, 255, 419

\bibitem[{{Boller} {et~al.}(1996){Boller}, {Brandt}, \& {Fink}}]{boller96}
{Boller}, T., {Brandt}, W.~N., \& {Fink}, H. 1996, \aap, 305, 53

\bibitem[{{Boroson}(2002)}]{boroson02}
{Boroson}, T.~A. 2002, \apj, 565, 78

\bibitem[{{Breeveld} \& {Puchnarewicz}(1998)}]{breeveld98}
{Breeveld}, A.~A. \& {Puchnarewicz}, E.~M. 1998, \mnras, 295, 568

\bibitem[{{Caccianiga} {et~al.}(2015){Caccianiga}, {Ant{\'o}n}, {Ballo},
  {Foschini}, {Maccacaro}, {Della Ceca}, {Severgnini}, {March{\~a}}, {Mateos},
  \& {Sani}}]{caccianiga15}
{Caccianiga}, A., {Ant{\'o}n}, S., {Ballo}, L., {et~al.} 2015, \mnras, 451,
  1795

\bibitem[{{Calderone} {et~al.}(2012){Calderone}, {Ghisellini}, {Colpi}, \&
  {Dotti}}]{calderone12}
{Calderone}, G., {Ghisellini}, G., {Colpi}, M., \& {Dotti}, M. 2012, \mnras,
  424, 3081

\bibitem[{{Calderone} {et~al.}(2013){Calderone}, {Ghisellini}, {Colpi}, \&
  {Dotti}}]{calderone13}
{Calderone}, G., {Ghisellini}, G., {Colpi}, M., \& {Dotti}, M. 2013, \mnras,
  431, 210

\bibitem[{{Calderone} {et~al.}(2016){Calderone}, {Nicastro}, {Ghisellini},
  {Dotti}, {Sbarrato}, {Shankar}, \& {Colpi}}]{calderone16}
{Calderone}, G., {Nicastro}, L., {Ghisellini}, G., {et~al.} 2016, ArXiv
  e-prints [\eprint[arXiv]{1612.01580}]

\bibitem[{{Collin} {et~al.}(2006){Collin}, {Kawaguchi}, {Peterson}, \&
  {Vestergaard}}]{collin06}
{Collin}, S., {Kawaguchi}, T., {Peterson}, B.~M., \& {Vestergaard}, M. 2006,
  \aap, 456, 75

\bibitem[{{D'Ammando} {et~al.}(2016){D'Ammando}, {Orienti}, {Finke}, {Larsson},
  {Giroletti}, \& {Raiteri}}]{dammando16}
{D'Ammando}, F., {Orienti}, M., {Finke}, J., {et~al.} 2016, Galaxies, 4, 11

\bibitem[{{D'Ammando} {et~al.}(2012){D'Ammando}, {Orienti}, {Finke}, {Raiteri},
  {Angelakis}, {Fuhrmann}, {Giroletti}, {Hovatta}, {Max-Moerbeck}, {Perkins},
  {Readhead}, {Richards}, {Stawarz}, \& {Donato}}]{dammando12}
{D'Ammando}, F., {Orienti}, M., {Finke}, J., {et~al.} 2012, \mnras, 426, 317

\bibitem[{{D'Ammando} {et~al.}(2015){D'Ammando}, {Orienti}, {Larsson}, \&
  {Giroletti}}]{dammando15}
{D'Ammando}, F., {Orienti}, M., {Larsson}, J., \& {Giroletti}, M. 2015, \mnras,
  452, 520

\bibitem[{{Decarli} {et~al.}(2008{\natexlab{a}}){Decarli}, {Dotti}, {Fontana},
  \& {Haardt}}]{decarli08b}
{Decarli}, R., {Dotti}, M., {Fontana}, M., \& {Haardt}, F. 2008{\natexlab{a}},
  \mnras, 386, L15

\bibitem[{{Decarli} {et~al.}(2011){Decarli}, {Dotti}, \& {Treves}}]{decarli11}
{Decarli}, R., {Dotti}, M., \& {Treves}, A. 2011, \mnras, 413, 39

\bibitem[{{Decarli} {et~al.}(2008{\natexlab{b}}){Decarli}, {Labita}, {Treves},
  \& {Falomo}}]{decarli08a}
{Decarli}, R., {Labita}, M., {Treves}, A., \& {Falomo}, R. 2008{\natexlab{b}},
  \mnras, 387, 1237

\bibitem[{{Fine} {et~al.}(2011){Fine}, {Jarvis}, \& {Mauch}}]{fine11}
{Fine}, S., {Jarvis}, M.~J., \& {Mauch}, T. 2011, \mnras, 412, 213

\bibitem[{{Foschini}(2011)}]{foschini11}
{Foschini}, L. 2011, in Narrow-Line Seyfert 1 Galaxies and their Place in the
  Universe, 24

\bibitem[{{Foschini}(2012)}]{foschini12}
{Foschini}, L. 2012, in American Institute of Physics Conference Series, Vol.
  1505, American Institute of Physics Conference Series, ed. F.~A. {Aharonian},
  W.~{Hofmann}, \& F.~M. {Rieger}, 574--577

\bibitem[{{Foschini}(2017)}]{foschini17}
{Foschini}, L. 2017, ArXiv e-prints [\eprint[arXiv]{1705.10166}]

\bibitem[{{Foschini} {et~al.}(2015){Foschini}, {Berton}, {Caccianiga}, {Ciroi},
  {Cracco}, {Peterson}, {Angelakis}, {Braito}, {Fuhrmann}, {Gallo}, {Grupe},
  {J{\"a}rvel{\"a}}, {Kaufmann}, {Komossa}, {Kovalev}, {L{\"a}hteenm{\"a}ki},
  {Lisakov}, {Lister}, {Mathur}, {Richards}, {Romano}, {Sievers},
  {Tagliaferri}, {Tammi}, {Tibolla}, {Tornikoski}, {Vercellone}, {La Mura},
  {Maraschi}, \& {Rafanelli}}]{foschini15}
{Foschini}, L., {Berton}, M., {Caccianiga}, A., {et~al.} 2015, \aap, 575, A13

\bibitem[{{Ghisellini} {et~al.}(2004){Ghisellini}, {Haardt}, \&
  {Matt}}]{ghisellini04}
{Ghisellini}, G., {Haardt}, F., \& {Matt}, G. 2004, \aap, 413, 535

\bibitem[{{Ghisellini} {et~al.}(2017){Ghisellini}, {Righi}, {Costamante}, \&
  {Tavecchio}}]{ghisellini17}
{Ghisellini}, G., {Righi}, C., {Costamante}, L., \& {Tavecchio}, F. 2017, ArXiv
  e-prints [\eprint[arXiv]{1702.02571}]

\bibitem[{{Ghisellini} \& {Sbarrato}(2016)}]{ghisellini16}
{Ghisellini}, G. \& {Sbarrato}, T. 2016, \mnras, 461, L21

\bibitem[{{Ghisellini} {et~al.}(2005){Ghisellini}, {Tavecchio}, \&
  {Chiaberge}}]{ghisellini05}
{Ghisellini}, G., {Tavecchio}, F., \& {Chiaberge}, M. 2005, \aap, 432, 401

\bibitem[{{Giommi} {et~al.}(2012){Giommi}, {Padovani}, {Polenta}, {Turriziani},
  {D'Elia}, \& {Piranomonte}}]{giommi12}
{Giommi}, P., {Padovani}, P., {Polenta}, G., {et~al.} 2012, \mnras, 420, 2899

\bibitem[{{Goodrich}(1989)}]{goodrich89}
{Goodrich}, R.~W. 1989, \apj, 342, 224

\bibitem[{{Green} {et~al.}(1993){Green}, {McHardy}, \& {Lehto}}]{green93}
{Green}, A.~R., {McHardy}, I.~M., \& {Lehto}, H.~J. 1993, \mnras, 265, 664

\bibitem[{{Greene} \& {Ho}(2005)}]{greene05}
{Greene}, J.~E. \& {Ho}, L.~C. 2005, \apj, 630, 122

\bibitem[{{Grier} {et~al.}(2017){Grier}, {Pancoast}, {Barth}, {Fausnaugh},
  {Brewer}, {Treu}, \& {Peterson}}]{grier17}
{Grier}, C.~J., {Pancoast}, A., {Barth}, A.~J., {et~al.} 2017, ArXiv e-prints
  [\eprint[arXiv]{1705.02346}]

\bibitem[{{Grupe} \& {Mathur}(2004)}]{grupe04}
{Grupe}, D. \& {Mathur}, S. 2004, \apjl, 606, L41

\bibitem[{{Hayashida}(2000)}]{hayashida00}
{Hayashida}, K. 2000, \nar, 44, 419

\bibitem[{{J{\"a}rvel{\"a}} {et~al.}(2015){J{\"a}rvel{\"a}},
  {L{\"a}hteenm{\"a}ki}, \& {Le{\'o}n-Tavares}}]{jarvela15}
{J{\"a}rvel{\"a}}, E., {L{\"a}hteenm{\"a}ki}, A., \& {Le{\'o}n-Tavares}, J.
  2015, \aap, 573, A76

\bibitem[{{Kaspi} {et~al.}(2005){Kaspi}, {Maoz}, {Netzer}, {Peterson},
  {Vestergaard}, \& {Jannuzi}}]{kaspi05}
{Kaspi}, S., {Maoz}, D., {Netzer}, H., {et~al.} 2005, \apj, 629, 61

\bibitem[{{Kaspi} {et~al.}(2000){Kaspi}, {Smith}, {Netzer}, {Maoz}, {Jannuzi},
  \& {Giveon}}]{kaspi00}
{Kaspi}, S., {Smith}, P.~S., {Netzer}, H., {et~al.} 2000, \apj, 533, 631

\bibitem[{{Kay} {et~al.}(1999){Kay}, {Magalh{\~a}es}, {Elizalde}, \&
  {Rodrigues}}]{kay99}
{Kay}, L.~E., {Magalh{\~a}es}, A.~M., {Elizalde}, F., \& {Rodrigues}, C. 1999,
  \apj, 518, 219

\bibitem[{{Labita} {et~al.}(2006){Labita}, {Treves}, {Falomo}, \&
  {Uslenghi}}]{labita06}
{Labita}, M., {Treves}, A., {Falomo}, R., \& {Uslenghi}, M. 2006, \mnras, 373,
  551

\bibitem[{{L{\"a}hteenm{\"a}ki} {et~al.}(2017){L{\"a}hteenm{\"a}ki},
  {J{\"a}rvel{\"a}}, {Hovatta}, {Tornikoski}, {Harrison}, {L{\'o}pez-Caniego},
  {Max-Moerbeck}, {Mingaliev}, {Pearson}, {Ramakrishnan}, {Readhead}, {Reeves},
  {Richards}, {Sotnikova}, \& {Tammi}}]{lahteenmaki17}
{L{\"a}hteenm{\"a}ki}, A., {J{\"a}rvel{\"a}}, E., {Hovatta}, T., {et~al.} 2017,
  ArXiv e-prints [\eprint[arXiv]{1703.10365}]

\bibitem[{{Miller} \& {Antonucci}(1983)}]{miller83}
{Miller}, J.~S. \& {Antonucci}, R.~R.~J. 1983, \apjl, 271, L7

\bibitem[{{Osterbrock} \& {Pogge}(1985)}]{osterbrock85}
{Osterbrock}, D.~E. \& {Pogge}, R.~W. 1985, \apj, 297, 166

\bibitem[{{Padovani}(2016)}]{padovani16}
{Padovani}, P. 2016, \aapr, 24, 13

\bibitem[{{Pancoast} {et~al.}(2014{\natexlab{a}}){Pancoast}, {Brewer}, \&
  {Treu}}]{pancoast14a}
{Pancoast}, A., {Brewer}, B.~J., \& {Treu}, T. 2014{\natexlab{a}}, \mnras, 445,
  3055

\bibitem[{{Pancoast} {et~al.}(2014{\natexlab{b}}){Pancoast}, {Brewer}, {Treu},
  {Park}, {Barth}, {Bentz}, \& {Woo}}]{pancoast14b}
{Pancoast}, A., {Brewer}, B.~J., {Treu}, T., {et~al.} 2014{\natexlab{b}},
  \mnras, 445, 3073

\bibitem[{{Rakshit} \& {Stalin}(2017)}]{rakshit17}
{Rakshit}, S. \& {Stalin}, C.~S. 2017, ArXiv e-prints
  [\eprint[arXiv]{1705.05123}]

\bibitem[{{Runnoe} {et~al.}(2013){Runnoe}, {Brotherton}, {Shang}, {Wills}, \&
  {DiPompeo}}]{runnoe13}
{Runnoe}, J.~C., {Brotherton}, M.~S., {Shang}, Z., {Wills}, B.~J., \&
  {DiPompeo}, M.~A. 2013, \mnras, 429, 135

\bibitem[{{Sani} {et~al.}(2010){Sani}, {Lutz}, {Risaliti}, {Netzer}, {Gallo},
  {Trakhtenbrot}, {Sturm}, \& {Boller}}]{sani10}
{Sani}, E., {Lutz}, D., {Risaliti}, G., {et~al.} 2010, \mnras, 403, 1246

\bibitem[{{Shakura} \& {Sunyaev}(1973)}]{ss73}
{Shakura}, N.~I. \& {Sunyaev}, R.~A. 1973, \aap, 24, 337

\bibitem[{{Shen} \& {Ho}(2014)}]{shen14}
{Shen}, Y. \& {Ho}, L.~C. 2014, \nat, 513, 210

\bibitem[{{Smith} {et~al.}(2004){Smith}, {Robinson}, {Alexander}, {Young},
  {Axon}, \& {Corbett}}]{smith04}
{Smith}, J.~E., {Robinson}, A., {Alexander}, D.~M., {et~al.} 2004, \mnras, 350,
  140

\bibitem[{{Smith} {et~al.}(2005){Smith}, {Robinson}, {Young}, {Axon}, \&
  {Corbett}}]{smith05}
{Smith}, J.~E., {Robinson}, A., {Young}, S., {Axon}, D.~J., \& {Corbett}, E.~A.
  2005, \mnras, 359, 846

\bibitem[{{V{\'e}ron-Cetty} {et~al.}(2001){V{\'e}ron-Cetty}, {V{\'e}ron}, \&
  {Gon{\c c}alves}}]{veron01}
{V{\'e}ron-Cetty}, M.-P., {V{\'e}ron}, P., \& {Gon{\c c}alves}, A.~C. 2001,
  \aap, 372, 730

\bibitem[{{Wills} \& {Browne}(1986)}]{wills86}
{Wills}, B.~J. \& {Browne}, I.~W.~A. 1986, \apj, 302, 56

\bibitem[{{Zhou} {et~al.}(2006){Zhou}, {Wang}, {Yuan}, {Lu}, {Dong}, {Wang}, \&
  {Lu}}]{zhou06}
{Zhou}, H., {Wang}, T., {Yuan}, W., {et~al.} 2006, \apjs, 166, 128

\end{thebibliography}

\end{document}